\documentclass[
reprint,
superscriptaddress,
 amsmath,amssymb,
 aps,
prl,
]{revtex4-2}

\usepackage{float}
\usepackage{graphicx}
\usepackage{dcolumn}
\usepackage{bm}
\usepackage{xcolor}
\usepackage{lipsum}  
\usepackage{hyperref}

\usepackage{lineno}
\begin{document}

\preprint{APS/PRL}

\title{Hosing of a long relativistic particle bunch in plasma}%

\author{T.~Nechaeva}
\email{tnechaev@mpp.mpg.de}
\affiliation{Max Planck Institute for Physics, 80805 Munich, Germany}

\author{L.~Verra}
\affiliation{CERN, 1211 Geneva 23, Switzerland}
\author{J.~Pucek}
\affiliation{Max Planck Institute for Physics, 80805 Munich, Germany}
\author{L.~Ranc}
\affiliation{Max Planck Institute for Physics, 80805 Munich, Germany}
\author{M.~Bergamaschi} 
\affiliation{Max Planck Institute for Physics, 80805 Munich, Germany}
\author{G.~Zevi Della Porta}
\affiliation{Max Planck Institute for Physics, 80805 Munich, Germany}
\affiliation{CERN, 1211 Geneva 23, Switzerland}
\author{P.~Muggli}
\affiliation{Max Planck Institute for Physics, 80805 Munich, Germany}


\collaboration{AWAKE Collaboration}
\noaffiliation
\author{R.~Agnello}
\affiliation{Ecole Polytechnique Federale de Lausanne (EPFL), Swiss Plasma Center (SPC), 1015 Lausanne, Switzerland}
\author{C.C.~Ahdida}
\affiliation{CERN, 1211 Geneva 23, Switzerland}
\author{C.~Amoedo}
\affiliation{CERN, 1211 Geneva 23, Switzerland}
\author{Y.~Andrebe}
\affiliation{Ecole Polytechnique Federale de Lausanne (EPFL), Swiss Plasma Center (SPC), 1015 Lausanne, Switzerland}
\author{O.~Apsimon}
\affiliation{University of Manchester M13 9PL, Manchester M13 9PL, United Kingdom}
\affiliation{Cockcroft Institute, Warrington WA4 4AD, United Kingdom}
\author{R.~Apsimon}
\affiliation{Cockcroft Institute, Warrington WA4 4AD, United Kingdom} 
\affiliation{Lancaster University, Lancaster LA1 4YB, United Kingdom}
\author{J.M.~Arnesano}
\affiliation{CERN, 1211 Geneva 23, Switzerland}
\author{V.~Bencini}
\affiliation{CERN, 1211 Geneva 23, Switzerland}
\affiliation{John Adams Institute, Oxford University, Oxford OX1 3RH, United Kingdom}
\author{P.~Blanchard}
\affiliation{Ecole Polytechnique Federale de Lausanne (EPFL), Swiss Plasma Center (SPC), 1015 Lausanne, Switzerland}
\author{P.N.~Burrows}
\affiliation{John Adams Institute, Oxford University, Oxford OX1 3RH, United Kingdom}
\author{B.~Buttensch{\"o}n}
\affiliation{Max Planck Institute for Plasma Physics, 17491 Greifswald, Germany}
\author{A.~Caldwell}
\affiliation{Max Planck Institute for Physics, 80805 Munich, Germany}
\author{M.~Chung}
\affiliation{UNIST, Ulsan 44919, Republic of Korea}
\author{D.A.~Cooke}
\affiliation{UCL, London WC1 6BT, United Kingdom}
\author{C.~Davut}
\affiliation{University of Manchester M13 9PL, Manchester M13 9PL, United Kingdom}
\affiliation{Cockcroft Institute, Warrington WA4 4AD, United Kingdom} 
\author{G.~Demeter}
\affiliation{Wigner Research Centre for Physics, 1121 Budapest, Hungary}
\author{A.C.~Dexter}
\affiliation{Cockcroft Institute, Warrington WA4 4AD, United Kingdom} 
\affiliation{Lancaster University, Lancaster LA1 4YB, United Kingdom}
\author{S.~Doebert}
\affiliation{CERN, 1211 Geneva 23, Switzerland}
\author{J.~Farmer}
\affiliation{Max Planck Institute for Physics, 80805 Munich, Germany}
\author{A.~Fasoli}
\affiliation{Ecole Polytechnique Federale de Lausanne (EPFL), Swiss Plasma Center (SPC), 1015 Lausanne, Switzerland}
\author{R.~Fonseca}
\affiliation{ISCTE - Instituto Universit\'{e}ario de Lisboa, 1049-001 Lisbon, Portugal}  
\affiliation{GoLP/Instituto de Plasmas e Fus\~{a}o Nuclear, Instituto Superior T\'{e}cnico, Universidade de Lisboa, 1049-001 Lisbon, Portugal}
\author{I.~Furno}
\affiliation{Ecole Polytechnique Federale de Lausanne (EPFL), Swiss Plasma Center (SPC), 1015 Lausanne, Switzerland}
\author{E.~Granados}
\affiliation{CERN, 1211 Geneva 23, Switzerland}
\author{M.~Granetzny}
\affiliation{University of Wisconsin, Madison, WI 53706, USA}
\author{T.~Graubner}
\affiliation{Philipps-Universit{\"a}t Marburg, 35032 Marburg, Germany}
\author{O.~Grulke}
\affiliation{Max Planck Institute for Plasma Physics, 17491 Greifswald, Germany}
\affiliation{Technical University of Denmark, 2800 Kgs. Lyngby, Denmark}
\author{E.~Gschwendtner}
\affiliation{CERN, 1211 Geneva 23, Switzerland}
\author{E.~Guran}
\affiliation{CERN, 1211 Geneva 23, Switzerland}
\author{J.~Henderson}
\affiliation{Cockcroft Institute, Warrington WA4 4AD, United Kingdom}
\affiliation{STFC/ASTeC, Daresbury Laboratory, Warrington WA4 4AD, United Kingdom}
\author{M.Á.~Kedves}
\affiliation{Wigner Research Centre for Physics, 1121 Budapest, Hungary}
\author{S.-Y.~Kim}
\affiliation{UNIST, Ulsan 44919, Republic of Korea}
\affiliation{CERN, 1211 Geneva 23, Switzerland}
\author{F.~Kraus}
\affiliation{Philipps-Universit{\"a}t Marburg, 35032 Marburg, Germany}
\author{M.~Krupa}
\affiliation{CERN, 1211 Geneva 23, Switzerland}
\author{T.~Lefevre}
\affiliation{CERN, 1211 Geneva 23, Switzerland}
\author{L.~Liang}
\affiliation{University of Manchester M13 9PL, Manchester M13 9PL, United Kingdom}
\affiliation{Cockcroft Institute, Warrington WA4 4AD, United Kingdom}
\author{S.~Liu}
\affiliation{TRIUMF, Vancouver, BC V6T 2A3, Canada}
\author{N.~Lopes}
\affiliation{GoLP/Instituto de Plasmas e Fus\~{a}o Nuclear, Instituto Superior T\'{e}cnico, Universidade de Lisboa, 1049-001 Lisbon, Portugal}
\author{K.~Lotov}
\affiliation{Budker Institute of Nuclear Physics SB RAS, 630090 Novosibirsk, Russia}
\affiliation{Novosibirsk State University, 630090 Novosibirsk , Russia}
\author{M.~Martinez~Calderon}
\affiliation{CERN, 1211 Geneva 23, Switzerland}
\author{S.~Mazzoni}
\affiliation{CERN, 1211 Geneva 23, Switzerland}
\author{K.~Moon}
\affiliation{UNIST, Ulsan 44919, Republic of Korea}
\author{P.I.~Morales~Guzm\'{a}n}
\affiliation{Max Planck Institute for Physics, 80805 Munich, Germany}
\author{M.~Moreira}
\affiliation{GoLP/Instituto de Plasmas e Fus\~{a}o Nuclear, Instituto Superior T\'{e}cnico, Universidade de Lisboa, 1049-001 Lisbon, Portugal}
\author{N.~Okhotnikov}
\affiliation{Budker Institute of Nuclear Physics SB RAS, 630090 Novosibirsk, Russia}
\affiliation{Novosibirsk State University, 630090 Novosibirsk , Russia}
\author{C.~Pakuza}
\affiliation{John Adams Institute, Oxford University, Oxford OX1 3RH, United Kingdom}
\author{F.~Pannell}
\affiliation{UCL, London WC1 6BT, United Kingdom}
\author{A.~Pardons}
\affiliation{CERN, 1211 Geneva 23, Switzerland}
\author{K.~Pepitone}
\affiliation{Angstrom Laboratory, Department of Physics and Astronomy, 752 37 Uppsala, Sweden}
\author{E.~Poimenidou}
\affiliation{CERN, 1211 Geneva 23, Switzerland}
\author{A.~Pukhov}
\affiliation{Heinrich-Heine-Universit{\"a}t D{\"u}sseldorf, 40225 D{\"u}sseldorf, Germany}
\affiliation{John Adams Institute, Oxford University, Oxford OX1 3RH, United Kingdom}
\author{S.~Rey}
\affiliation{CERN, 1211 Geneva 23, Switzerland}
\author{R.~Rossel}
\affiliation{CERN, 1211 Geneva 23, Switzerland}
\author{H.~Saberi}
\affiliation{University of Manchester M13 9PL, Manchester M13 9PL, United Kingdom}
\affiliation{Cockcroft Institute, Warrington WA4 4AD, United Kingdom}
\author{O.~Schmitz}
\affiliation{University of Wisconsin, Madison, WI 53706, USA}
\author{E.~Senes}
\affiliation{CERN, 1211 Geneva 23, Switzerland}
\author{F.~Silva}
\affiliation{INESC-ID, Instituto Superior Técnico, Universidade de Lisboa, 1049-001 Lisbon, Portugal}
\author{L.~Silva}
\affiliation{GoLP/Instituto de Plasmas e Fus\~{a}o Nuclear, Instituto Superior T\'{e}cnico, Universidade de Lisboa, 1049-001 Lisbon, Portugal}
\author{B.~Spear}
\affiliation{John Adams Institute, Oxford University, Oxford OX1 3RH, United Kingdom}
\author{C.~Stollberg}
\affiliation{Ecole Polytechnique Federale de Lausanne (EPFL), Swiss Plasma Center (SPC), 1015 Lausanne, Switzerland}
\author{A.~Sublet}
\affiliation{CERN, 1211 Geneva 23, Switzerland}
\author{C.~Swain}
\affiliation{Cockcroft Institute, Warrington WA4 4AD, United Kingdom}
\affiliation{University of Liverpool, Liverpool L69 7ZE, United Kingdom}
\author{A.~Topaloudis}
\affiliation{CERN, 1211 Geneva 23, Switzerland}
\author{N.~Torrado}
\affiliation{GoLP/Instituto de Plasmas e Fus\~{a}o Nuclear, Instituto Superior T\'{e}cnico, Universidade de Lisboa, 1049-001 Lisbon, Portugal}
\affiliation{CERN, 1211 Geneva 23, Switzerland}
\author{M.~Turner}
\affiliation{CERN, 1211 Geneva 23, Switzerland}
\author{F.~Velotti}
\affiliation{CERN, 1211 Geneva 23, Switzerland}
\author{V.~Verzilov}
\affiliation{TRIUMF, Vancouver, BC V6T 2A3, Canada}
\author{J.~Vieira}
\affiliation{GoLP/Instituto de Plasmas e Fus\~{a}o Nuclear, Instituto Superior T\'{e}cnico, Universidade de Lisboa, 1049-001 Lisbon, Portugal}
\author{C.~Welsch}
\affiliation{Cockcroft Institute, Warrington WA4 4AD, United Kingdom}
\affiliation{University of Liverpool, Liverpool L69 7ZE, United Kingdom}
\author{M.~Wendt}
\affiliation{CERN, 1211 Geneva 23, Switzerland}
\author{M.~Wing}
\affiliation{UCL, London WC1 6BT, United Kingdom}
\author{J.~Wolfenden}
\affiliation{Cockcroft Institute, Warrington WA4 4AD, United Kingdom}
\affiliation{University of Liverpool, Liverpool L69 7ZE, United Kingdom}
\author{B.~Woolley}
\affiliation{CERN, 1211 Geneva 23, Switzerland}
\author{G.~Xia}
\affiliation{Cockcroft Institute, Warrington WA4 4AD, United Kingdom}
\affiliation{University of Manchester M13 9PL, Manchester M13 9PL, United Kingdom}
\author{V.~Yarygova}
\affiliation{Budker Institute of Nuclear Physics SB RAS, 630090 Novosibirsk, Russia}
\affiliation{Novosibirsk State University, 630090 Novosibirsk , Russia}
\author{M.~Zepp}
\affiliation{University of Wisconsin, Madison, WI 53706, USA}


\begin{abstract}

Experimental results show that hosing of a long particle bunch in plasma can be induced by wakefields driven by a short, misaligned preceding bunch. Hosing develops in the plane of misalignment, self-modulation in the perpendicular plane, at frequencies close to the plasma electron frequency, and are reproducible. Development of hosing depends on misalignment direction, its growth on misalignment extent and on proton bunch charge. Results have the main characteristics of a theoretical model, are relevant to other plasma-based accelerators and represent the first characterization of hosing.

\end{abstract}

\maketitle


\section{\label{sec:level1} Introduction}

Hosing of a charged particle bunch in plasma is a fundamental mode of interaction and instability of such a system \cite{whittum}. %
Studying hosing is important because it could impose a limit on the distance a bunch can propagate in plasma. %
This is the case, e.g., when using a pre-formed plasma to guide a beam through the atmosphere (\cite{resh}, resistive hosing). %
Hosing is posited to be a limit for propagation of both witness \cite{witnpluscollider} and drive bunches \cite{sim1,hsupp2} in plasma-based accelerators. %
These accelerators are of interest and importance because they can operate at much higher accelerating gradients (1 -- 100 GeV/m \cite{pwfa2007,pwfa2019}) than conventional, RF-based accelerators ($< 1$ GeV/m \cite{rfcav2003, rfcav2016}).
Development of hosing would disrupt driving wakefields and the acceleration process, thus, the quality of the witness bunch. %
Choosing accelerator parameters to avoid hosing might limit the efficiency of the energy transfer process \cite{witnpluscollider}, a crucial parameter, e.g., for collider applications. %
The disruptive nature of hosing motivates studying its mitigation \cite{hsupp1,hsupp2,hsupp3,hsupp4}. %

Hosing of charged particle bunches propagating in plasma is well described by theory \cite{whittum,resh,theor1,theor2,schroeder} and numerical simulations \cite{sim1}. %
While we focus on the case of particle bunches, laser pulses propagating in plasma are also subject to this instability \cite{lashos1,lashos2}, though the underlying physics is quite different.%

Hosing occurs when the centroid position of a particle bunch couples to that of the focusing force (i.e., position where the axisymmetric force is zero) exerted by plasma \cite{whittum}, or to that of the wakefields \cite{hsupp1}. %
With a long bunch, the axisymmetric wakefields force leads to self-modulation (SM), which transforms the bunch into a train of microbunches \cite{kumarsmi,adliprl}.
Non-axisymmetric coupling results in transverse oscillation of the two centroid positions. 
When viewed in the laboratory reference frame, growth occurs because the displacement of each successive transverse slice of the bunch depends on that of all previous slices, through the focusing force sustained by oscillating plasma electrons. %
Growth thus occurs along the bunch and plasma. %
Hosing can initiate from global misalignment of the bunch with respect to a pre-existing focusing structure, notably in the case of a pre-formed focusing channel \cite{whittum}. Alternatively, this process could start from variations in the local bunch centroid position, e.g., when the focusing force is driven by the bunch itself \cite{schroeder,hsupp4}. %

Transverse displacement of the centroid position of an electron bunch or a laser pulse was observed in a number of plasma wakefield experiments and was attributed to the occurrence of hosing \cite{hismi, lashosexp}. However, very few fundamental characteristics of hosing (other than its possible occurrence) were deduced.

In this \textit{Letter}, we show that hosing of a long, relativistic proton bunch propagating in an over-dense plasma can be induced, and thus observed in a reproducible way. 
Hosing is induced by relative misalignment between the trajectory of a short electron bunch, hence the wakefields it drives, and that of the trailing proton bunch \cite{myhosproc}. Hosing and SM develop simultaneously, in perpendicular planes.
With no electron bunch, SM instability without hosing occurs \cite{fabprl}, showing that wakefields induce hosing.
The electron bunch drives wakefields at the plasma electron frequency $f_{pe} = \omega_{pe}/2\pi$ \cite{refomegape}, therefore the frequency of hosing is close to that of SM, both close to $f_{pe}$. %
The development of hosing depends on the direction of misalignment. %
For misalignment extents $\Delta x > 0.5\ c/\omega_{pe}$ ($c/\omega_{pe}$ -- cold plasma skin depth), the amplitude of hosing decreases with increasing $\Delta x$.
When $\Delta x > 2.5\ c/\omega_{pe}$, SM develops as an instability without hosing. 
The amplitude of hosing increases with larger proton bunch charge $Q_p$ (fixed $\Delta x$). %
We find good general agreement between hosing observed and a theoretical model \cite{schroeder}, despite differences between assumptions of the model and the experimental conditions. 
Finally, we note that SM grows because each microbunch drives its own wakefields, which, in the plane of SM, add to those driven by previous microbunches. %
These microbunches therefore drive wakefields in the plane of hosing as well, i.e., their off-axis wakefields add to those of the previous off-axis microbunches and thus lead to the growth of hosing. %
The $x_{c}$ oscillation we observe is therefore unambiguous evidence of the growth of hosing, as opposed to a betatron oscillation in the growing fields of SM. %

The results were obtained in the AWAKE experiment \cite{readiness}. Previously, hosing was observed only at low plasma densities ($<0.5\times10^{14}$ $\text{cm}^{-3}$) when SM was not seeded \cite{mathiasphd}, and has not been a limitation for acceleration experiments performed at higher densities. %
Numerical simulation results show that seeding SM with a relativistic ionization front suppresses hosing \cite{hsupp1}. %
However, seeding of SM with an electron bunch may be required in future experiments using, e.g., a pre-formed plasma \cite{patricrun2}. %
The results of the study presented here are thus an essential first step in identifying mechanisms that can seed hosing, and in understanding its development. %
Futher studies will have to determine tolerances in parameter space for SM seeded by the electron bunch to dominate over hosing.

We note that since the bunch density is lower than the plasma electron density, and the wakefields amplitude never reaches the wavebreaking amplitude, one can expect the evolution of hosing and SM to be essentially independent of the charge sign of the long bunch.
In addition, misalignment between drive and witness bunches in plasma-based accelerators using short drivers is also a seed for hosing. 
\section{\label{sec:level2} Experimental setup}

\begin{figure}[th!]
\centering
\includegraphics[width=1.0\columnwidth]{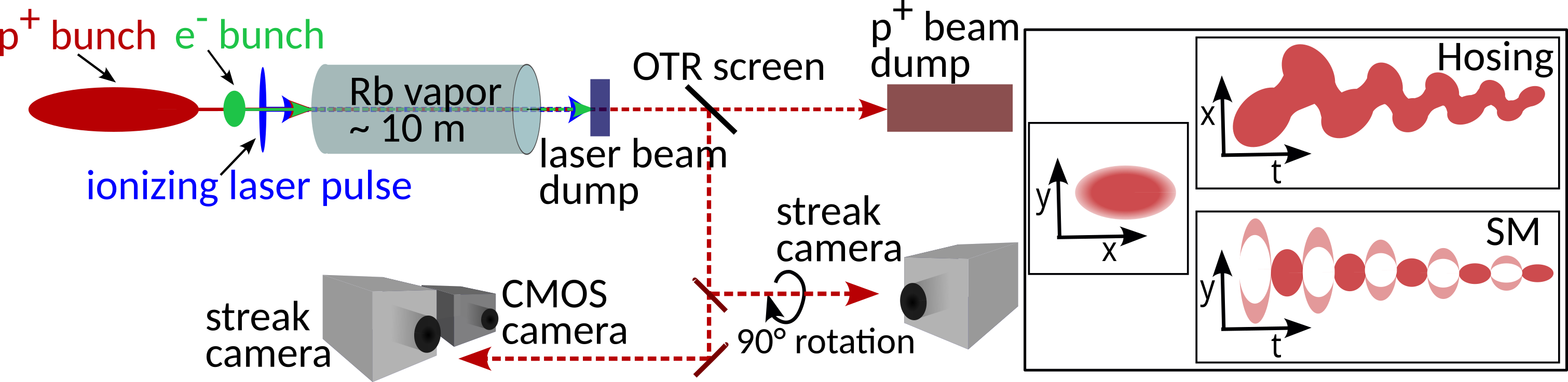}
\caption{\label{fig:expsetup} Schematic of the experimental setup (not to scale) showing main components used for the measurements. Inset: sketch of the images recorded: time-integrated (x, y), time-resolved (x, t), (y, t) images when hosing and SM develop simultaneously.}
\end{figure}

The proton bunch from the CERN Super Proton Synchrotron has an energy of 400 GeV per particle and an approximately Gaussian temporal distribution with root mean square (rms) duration $\sigma_{t} \approx 220$ ps. It has a round Gaussian rms transverse waist size $\sigma_{r0} \approx$ 0.2 mm at the entrance of a 10 m-long vapor source \cite{vaps} (Fig. \ref{fig:expsetup}). The source contains Rubidium vapor with uniform temperature, and therefore density \cite{rbdens} $n_{Rb}$, adjustable in the $(0.5-10)\times 10^{14}\ \text{cm}^{-3}$ range. The vapor, and, indirectly, the plasma density is measured to better than 0.5\% accuracy \cite{densmeas}. A Ti:Sapphire laser generates a $\sim 120$ fs-long, $\sim 110$ mJ pulse. 
This pulse provides full, single ionization of the Rubidium vapor, creating a plasma column of $\sim 1$ mm radius. 

The laser pulse propagates 620 ps ($\sim 2.8\sigma_{t}$) ahead of the proton bunch longitudinal center and is aligned on its axis, thus, it does not seed SM \cite{fabprl} or induce hosing. 
A 18.9 MeV, $\sim 225$ pC, $\sim 4$ ps-long electron bunch \cite{eaccel,livprl} placed 600 ps ahead of the proton bunch center, i.e., in plasma, 20 ps behind the laser pulse, drives seed wakefields. 

Protons pass through an aluminum-coated silicon wafer located 3.5 m downstream of the plasma exit, where they emit optical transition radiation (OTR). 
The OTR is transported, split and imaged onto the entrance slit of two streak cameras providing time-resolved images of the bunch charge density distribution \cite{karlstreak}. 
A $90^{\circ}$ spatial rotation is applied to one of the OTR signals, thus the images are in perpendicular planes \textit{(x, t)} and  \textit{(y, t)} (inset, Fig. \ref{fig:expsetup}). 
A CMOS camera yields time-integrated bunch charge distribution on the same OTR screen. 
This setup is crucial for detecting simultaneous occurrence of hosing and SM. 
The spatial resolution of the optical system is $\sim 0.18$ mm \cite{tatipac}. The temporal resolution of the streak cameras is $\sim 1$ ps \cite{tatipac,annambunch}, sufficient for the measurements presented here.

To circumvent the 5 ps rms jitter of the triggering system, thus, to determine the precise timing of hosing and SM along the bunch, we use a bleed-through of the ionizing laser pulse from a mirror. 
This pulse is synchronized with the main pulse and with the electron bunch. It is imaged onto the streak cameras and serves as a timing reference that allows temporal alignment of images at the sub-ps level \cite{fabmarker}. This is essential to demonstrate the reproducibility of the observed phenomena, as shown later.

\section{\label{sec:level3} Experimental results}

\begin{figure}[ht]
\includegraphics*[width=1\columnwidth]{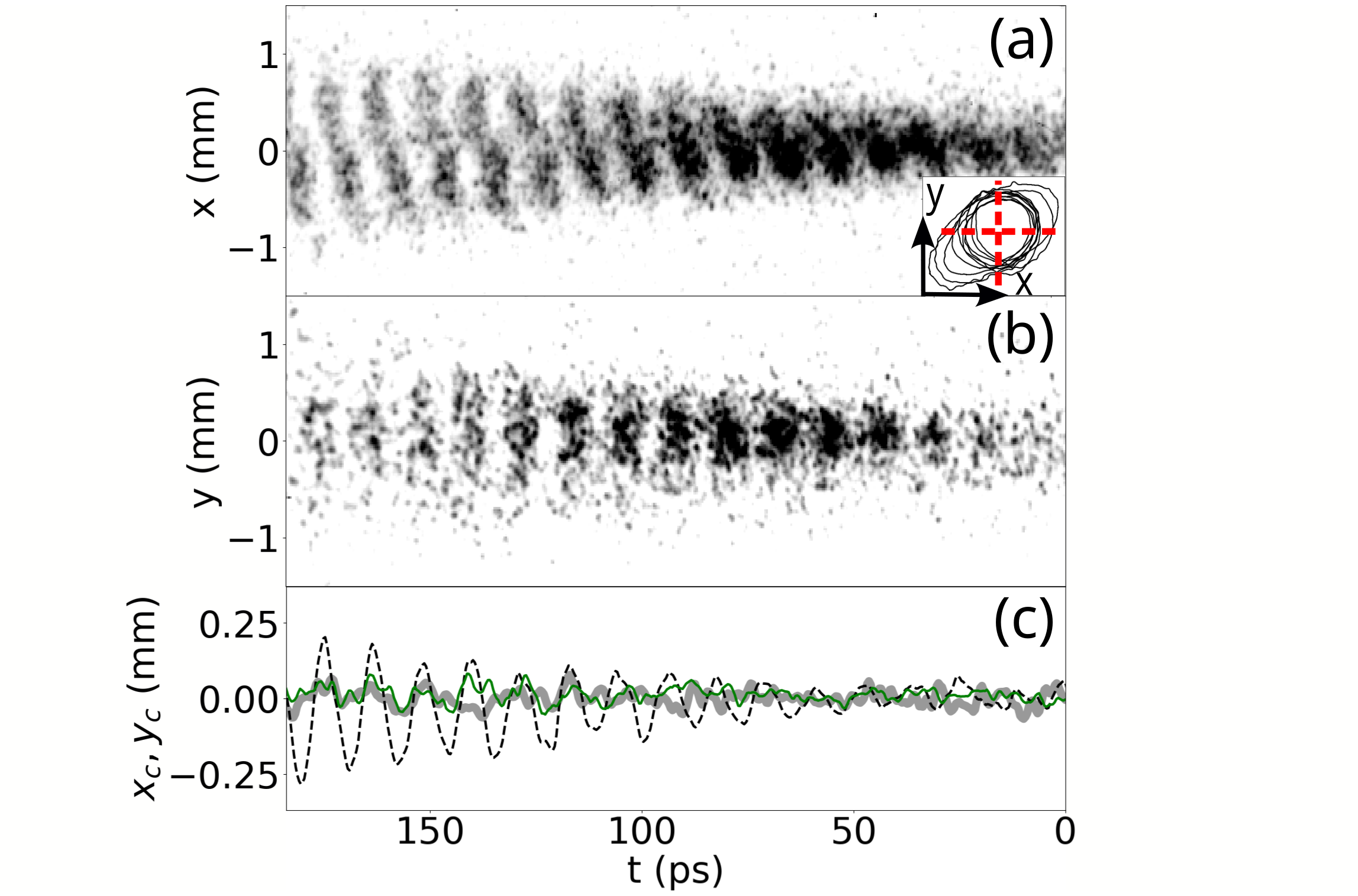}
\caption{\label{fig:fig2} Time-resolved images of the proton bunch ($n_{pe}$ = $0.96\times 10^{14}$ $\text{cm}^{-3}$, $Q_{p}$ = (14.9 $\pm$ 0.1) nC, $\Delta x$ = (0.95 $\pm$ 0.16) $c/\omega_{pe}$) (a) the x-plane -- hosing, (b) the y-plane -- SM. Images -- averages of ten consecutive single-events recorded simultaneously, same color scale. Inset (a): single-event 3$\sigma$ contours of time-integrated proton bunch charge distribution for the same ten events. Dashed red lines: position of the slit of the streak cameras. Elongation of the distribution indicates the plane of hosing. (c) Centroid position along the bunch in vacuum (grey line), undergoes SM (solid green line) and hosing (dashed black line).}
\end{figure}

We first introduce experimental observation of hosing and SM occurring simultaneously. The proton bunch has a charge $Q_{p}$ = (14.9 $\pm$ 0.1) nC and initially (in vacuum) a continuous charge density distribution. In order to induce hosing, we misalign the trajectory of the preceding electron bunch by $\Delta x$ = (0.95 $\pm$ 0.16) $c/\omega_{pe}$ (at $n_{pe}$ = $0.96\times 10^{14}$ $\text{cm}^{-3}$, i.e., $c/\omega_{pe} = 0.52$ mm) with respect to the proton bunch propagation axis, in the x-direction of the streak camera coordinate system. The $\Delta x$ is an average misalignment extent value with error bar representing the position jitter of both bunches summed in quadrature. 
Time-resolved images (Fig. \ref{fig:fig2}) show typical distributions corresponding to the simultaneous occurrence of hosing in the plane of misalignment (x, Fig. \ref{fig:fig2}(a)) and of SM in the perpendicular plane (y, Fig. \ref{fig:fig2}(b)).
We define $t=0$ as 279 ps ($\sim 1.3\sigma_t$) ahead of the longitudinal center of the proton bunch, propagating to the right.
Images aligned in time are averaged over ten consecutive events. 
The average images show clear ps-scale features of hosing, i.e., oscillation of centroid position, and of SM, i.e., microbunches \cite{fabprl,livprl}. This indicates that both processes are reproducible, and thus confirms that both are induced by the initial wakefields of the electron bunch \cite{livprl,fabprl}. 

The centroid position of the bunch in the SM plane (Fig. \ref{fig:fig2}(c), $y_{c}$, solid green line), and when propagating in vacuum (grey line, image not shown), remains close to the bunch axis (within $\pm$ 0.05 mm) and does not exhibit any periodic pattern. On the contrary, it clearly oscillates with growing amplitude (up to 0.28 mm) in the plane where hosing occurs ($x_{c}$, dashed black line). A $1.2 \times 0.07$ [ps, mm] median filter was applied to the time-resolved data to obtain smoother curves.  

We determine the frequencies of hosing ($f_{H}$) and SM ($f_{SM}$) by performing a discrete Fourier transform (DFT) of $x_{c} (t)$ for hosing and of the on-axis longitudinal (time) profile obtained within the proton bunch core radius for SM (not shown) \cite{karlstreak}. 
With $n_{pe}$ = 0.96$\times 10^{14}$ $\text{cm}^{-3}$ (Fig. \ref{fig:fig2}), i.e., $f_{pe}$ = (87.99 $\pm$ 0.18) GHz, we obtain $f_{H}$ = (86.76 $\pm$ 1.53) GHz and $f_{SM}$ = (86.27 $\pm$ 1.54) GHz, with $n_{pe}$ = 2.03$\times 10^{14}$ $\text{cm}^{-3}$ (curves not shown), i.e., $f_{pe}$ = (127.80 $\pm$ 0.26) GHz, $f_{H}$ = (125.31 $\pm$ 1.56) GHz and $f_{SM}$ = (125.41 $\pm$ 1.26) GHz. At both densities $f_{H}$ $\approx$ $f_{SM}$ $\approx$ $f_{pe}$.
This is expected, as the electron bunch drives initial wakefields at $f_{pe}$.

\begin{figure}[ht]
\centering
\includegraphics*[width=1.0\columnwidth]{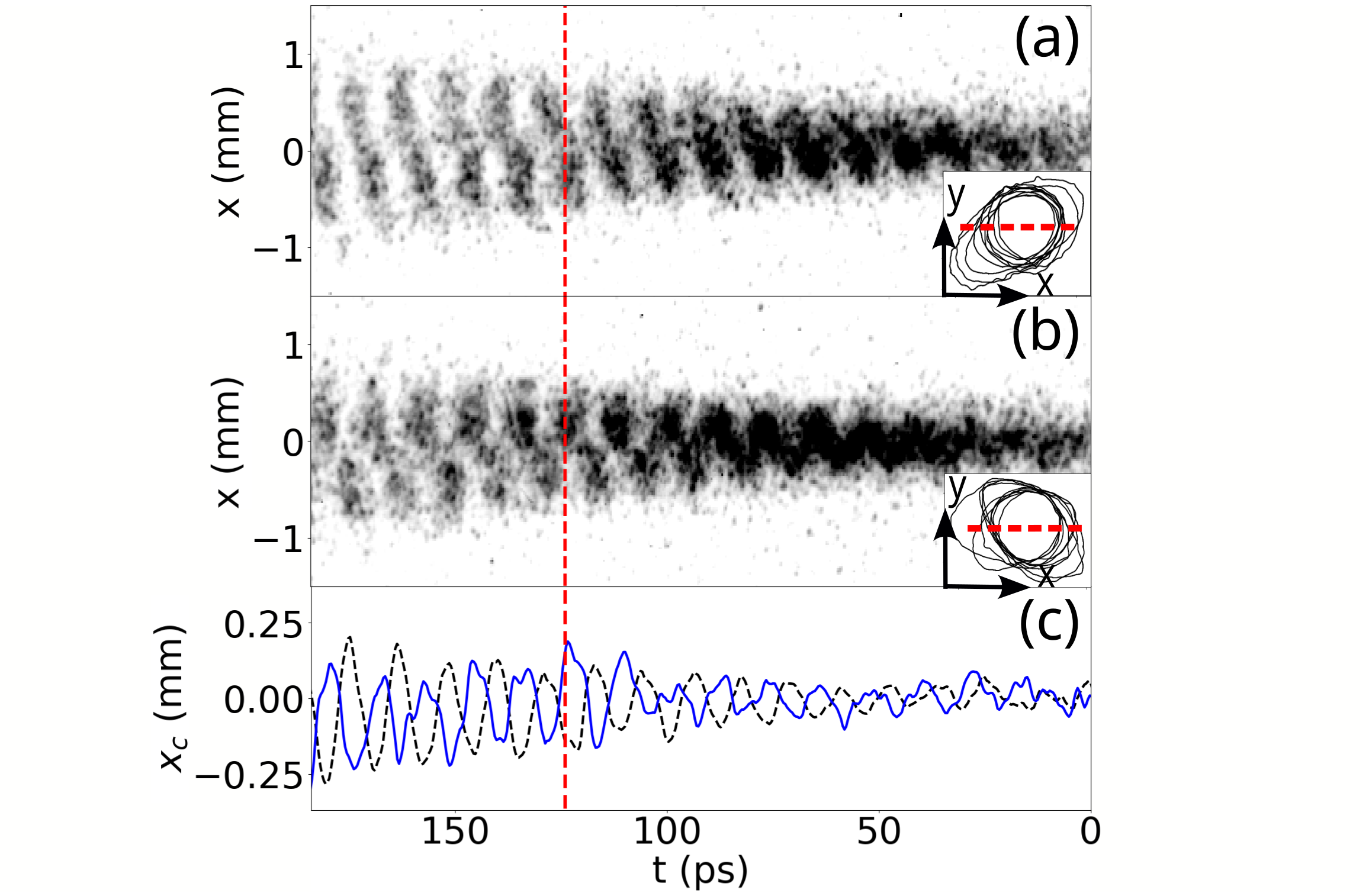}
\caption{\label{fig:fig3} Time-resolved images ($n_{pe}$ = $0.96\times 10^{14}\ \text{cm}^{-3}$, $Q_{p}$ = (14.9 $\pm$ 0.1) nC): (a) $\Delta x = (0.95 \pm 0.16)\ c/\omega_{pe}$, (b) $\Delta x = (-0.93 \pm 0.18)\ c/\omega_{pe}$. Images have the same color scale. Reflection of $x_c$ oscillation visible, e.g., at $t \approx$ 124 ps (dashed red line). (c) $x_c$ oscillation curves: dashed black line -- top image, solid blue line -- bottom image. Insets (a), (b): single-event $3\sigma$ contours of time-integrated bunch charge distribution.}
\end{figure}

Figure 3 shows that, when reversing the misalignment direction ($+\Delta x \rightarrow -\Delta x$), hence, the direction of the non-axisymmetric wakefield force acting on each slice of the bunch, the $x_c$ oscillation is reflected with respect to the bunch propagation axis.
This reflection is clearly visible in Figs. \ref{fig:fig3}(a) (same as Fig. \ref{fig:fig2}(a)), (b), and on the corresponding $x_{c}(t)$ curves (Fig. \ref{fig:fig3}(c), e.g., at $t \approx$ 124 ps, dashed red line).

Alignment in position and angle between the two particle beams with significantly different parameters is the main challenge in these experiments. 
Time-integrated images of the proton bunch transverse charge distribution are shown in insets of Fig. \ref{fig:fig2}(a) and Figs. \ref{fig:fig3}(a), (b) as 3$\sigma$ contours of the distribution of single events. These contours exhibit elongation expected when hosing occurs. %
They show that the direction of elongation is in the general direction of misalignment, i.e., x-direction, but with an angle of $\sim 27.5^{\circ}$ with respect to it. 
When $+\Delta x \rightarrow -\Delta x$, the angle is reflected from +27.5$^\circ$ to $180^\circ -26.2^\circ$.
This indicates a possible angular misalignment between the trajectories of the two bunches. %
This misalignment might also be the cause of the different amplitudes of the $x_c$ oscillation curves of Fig. \ref{fig:fig3}, since the amplitude depends on $\Delta x$ (see below).
The accuracy of the available diagnostics was not sufficient to correct this misalignment. %
Simulation results \cite{mathiasphd} indicate that time-resolved images nevertheless retain the main characteristics of both SM and hosing (as in Fig. \ref{fig:fig2}), even when the observation plane is different from the misalignment plane by angles similar to those observed in the experiment. %

\begin{figure}[ht]
\includegraphics*[width=1.0\columnwidth]{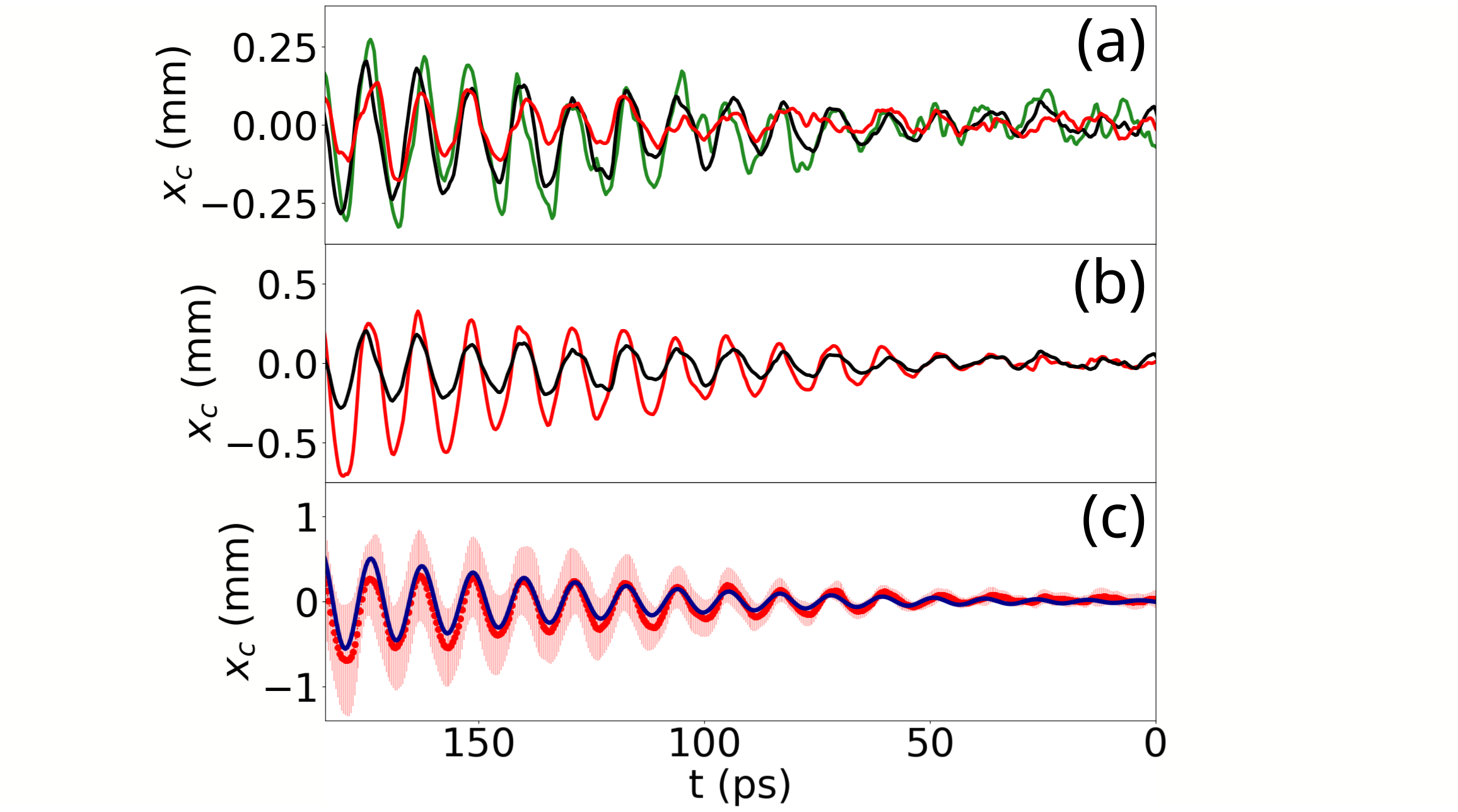}
\caption{\label{fig:fig4} 
Proton bunch centroid position $x_{c} (t)$ ($n_{pe}$ = $0.96\times 10^{14}\ \text{cm}^{-3}$). 
(a) $Q_{p} = (14.9 \pm 0.1)$ nC, $\Delta x = (0.53 \pm 0.15)\ c/\omega_{pe}$ -- green line, $\Delta x = (0.95 \pm 0.16)\ c/\omega_{pe}$ -- black line, $\Delta x = (1.47 \pm 0.16)\ c/\omega_{pe}$ -- red line. 
(b) $\Delta x \approx 1\ c/\omega_{pe}$. $Q_{p} = (46.5 \pm 0.6)$ nC ($n_{b0} = (7.0 \pm 0.9)\times 10^{12}\ \text{cm}^{-3}$) -- red line, $Q_{p} = (14.9 \pm 0.1)$ nC ($n_{b0} = (4.3 \pm 0.2)\times 10^{12}\ \text{cm}^{-3}$) -- black line.
(c) same as red line of (b) with rms variation of the data -- red bars and result of the fit of Eq. (1) -- blue line. $R^{2}$ = 0.84. 
}
\end{figure}

Numerical simulation results \cite{kookjin} show that the amplitude of the wakefields, driven by an electron bunch with parameters similar to those of the experiment, as a function of distance from the bunch axis peaks before $\Delta x = 0.5\ c/\omega_{pe}$ and then monotonically decreases.
The effect of these wakefields on the proton bunch centroid therefore depends on $\Delta x$ (other parameters kept constant).
Figure \ref{fig:fig4}(a) shows that, as expected, the amplitude of $x_c$ oscillation decreases as $\Delta x$ increases. The amplitude, measured at $t \approx 163$ ps, is  $x_c$ [$\Delta x \approx 0.5\ c/\omega_{pe}$]  $\approx 0.198$ mm (green line), $x_c$ [$\Delta x \approx 1.0\ c/\omega_{pe}$] $\approx 0.173$ mm (black line) and $x_c$ [$\Delta x \approx 1.5\ c/\omega_{pe}$] $\approx 0.108$ mm (red line).
When $\Delta x$ is sufficiently large ($> 2.5\ c/\omega_{pe}$, not shown) the wakefields effect is not strong enough to seed either hosing, that is not observed, or SM, that develops as an instability.

Theory (e.g., \cite{schroeder}) suggests that the number of exponentiations of hosing $N_h$, or the growth rate, increases with bunch density $n_{b0}$, i.e., with $Q_p$ (see Eq. (2) below). Results show that, with $Q_{p} \approx$ 46.5 nC ($n_{b0} \approx 7\times 10^{12}\ \text{cm}^{-3}$), the amplitude of hosing is on average 2.1 times higher (Fig. \ref{fig:fig4}(b), red line) than with $Q_{p} \approx$ 14.9 nC ($n_{b0} \approx 4.3\times 10^{12}\ \text{cm}^{-3}$, black line). We note that $n_{b0}$ varies less than $Q_{p}$ due to the change in the transverse emittance and hence transverse size of the bunch at the plasma entrance. Also, the measurement of $x_c$ is performed not in plasma (as in \cite{schroeder}), but after 3.5 m of propagation from the plasma exit to the OTR screen. Therefore the values of $x_c$ and $N_h$ that we calculate overestimate those in plasma.

The development of hosing, with or without presence of SM, in the long-beam early-time regime was considered theoretically \cite{schroeder}. 
In that study, the growth of the two processes starts from imposed initial centroid position (hosing) and radius (SM) perturbations of the bunch. 
Other assumptions are made for the derivation (see Supplemental Material \cite{suppref}). \nocite{eseedkookjin}

The asymptotic solution for the centroid position $x_c$ of the bunch with non-evolving radius is given by Eq. (10) in \cite{schroeder}. 
We convert the co-moving variable $\zeta$ of \cite{schroeder} to time $t$ for the analysis of the experimental results. 
We additionally introduce $t_0$ as the time when the amplitude growth starts ($t_0 = 0$ in \cite{schroeder}), since it is unknown in the experiment.
We rewrite Eqs. (10) and (11) of \cite{schroeder} as

\begin{equation}
	x_{c} = \delta_{c}\biggr[ \frac{3^{1/4}}{(8\pi^{1/2})}   \biggr] \frac{e^{N_{h}}}{N_{h}^{1/2}}\cos \biggl(\frac{\pi}{12} - \omega_{pe}(t - t_{0})-\frac{N_{h}}{\sqrt{3}} \biggl) 
\end{equation}
and
\begin{equation}	
N_{h} = \frac{3^{3/2}}{4} \biggl(\mu  \frac{m_{e}}{m_{b}}\frac{n_{b0}}{n_{pe}}\frac{1}{2\gamma} \biggl(\frac{\omega_{pe}}{c} \biggl)^{3}c(t - t_{0})z^{2} \biggl)^{1/3}.
\end{equation}

Here, $\mu$ represents the plasma return current, $c$ -- the speed of light, $m_b$ and $\gamma$ -- the mass and relativistic factor of the proton, and $z$ -- the distance along the plasma. 
The evolution of the amplitude of $x_c$ oscillation along the bunch is determined by $N_h$ as a function of $(t-t_0)$ at a given $z$ through the $\frac{e^{N_h}}{N_h^{1/2}}$ term.
The growth of oscillation starts from the initial periodic perturbation with amplitude $\delta_{c}$ (at $z=0$).  

Some of the assumptions used to obtain Eqs. (1) and (2) are not verified in the experiments (see Supplemental Material \cite{suppref}). Nevertheless, as the derivation was performed for similar conditions, we use Eq. (1) to determine whether the experimental results retain some of the characteristics highlighted in \cite{schroeder}. We perform a nonlinear least squares fit of this equation to the experimental $x_c$ curves, with initial free parameters $t_0$ and $\delta_{c}$. 
Figure \ref{fig:fig4}(c) shows that the data of Fig. \ref{fig:fig4}(b) ($Q_p \approx 46.5$ nC, red line) preserves the main characteristics of the model (blue line, fit, goodness $R^2=0.84$), i.e., $x_c$ oscillation at $f_H \approx f_{pe}$ growing along the bunch.
The result of the fit is well within the rms variations observed over ten events. %

We determine the growth of the amplitude of hosing from the experimental value of $x_c$ oscillation peak at $t \approx 163$ ps and $\delta_c$ obtained from the fit. With $x_c$ [$t \approx 163$ ps] $\approx 0.306$ mm and $\delta_c \approx 11.6\ \mu$m (Fig. \ref{fig:fig4}(c)), $x_c [t\approx 163\ \text{ps}]/\delta_c \approx 26.3$ ($N_h = 5.46$). 
Similar growth of the wakefields amplitude in case of SM was shown in \cite{marl}. This is expected, since theory suggests that hosing and SM have similar growth rates (e.g., \cite{schroeder}).
The results of the fit to all the data shown here and more detailed analysis can be found in the Supplemental Material \cite{suppref}.

Whilst all data presented shows the main features of hosing, when increasing $Q_p$, we observe a clear asymmetry of the $x_c$ oscillation with respect to the bunch axis (Fig. \ref{fig:fig4}(b), red line). 
Reference \cite{schroeder} shows that coupling between hosing and SM developing simultaneously (as in the experiment) generates such an asymmetry. The strength of the coupling and the resulting asymmetry depends on, e.g., initial seed amplitudes for hosing and SM. These parameters are not measured in the experiment and differ (wakefields driven by the electron bunch) from the ones in the theory (initial centroid and envelope perturbations). Quantitative comparison is therefore elusive.

\section{\label{sec:level4} Summary}

We observe hosing of a long proton bunch induced by the misalignment of the initial wakefields driven by a short electron bunch. Experimental results show a clear periodic centroid position oscillation that grows along the bunch and plasma, typical of hosing. Hosing occurs in the plane of misalignment, SM simultaneously in the perpendicular plane. Hosing and SM are induced by the same initial wakefields, therefore both processes are reproducible. Their frequencies are close to the plasma electron frequency. 
When reversing the misalignment direction, the centroid position oscillation is reflected with respect to the bunch propagation axis.
Its amplitude increases with proton bunch charge and tends to decrease with larger misalignment extents, as expected from theory and simulations findings. For misalignment extent larger than $2.5\ c/\omega_{pe}$ no hosing is observed, and SM develops as an instability.
The observed centroid position oscillation follows a theoretical model \cite{schroeder}.

Results show that misalignment of the initial wakefields induces hosing and has to be avoided in plasma-based accelerators. Studies of tolerance of the system to misalignment have to be conducted, especially when seeding SM with an electron bunch \cite{livprl}. 

\begin{acknowledgments}
\section{Acknowledgments}
This work was supported in parts by STFC (AWAKE-UK, Cockcroft Institute core, John Adams Institute core, and UCL consolidated grants), United Kingdom, the National Research
Foundation of Korea (Nos. NRF-2016R1A5A1013277 and NRF-2020R1A2C1010835). 
M. Wing acknowledges the support of DESY, Hamburg. 
Support of the Wigner Datacenter Cloud facility through
the Awakelaser project is acknowledged.
TRIUMF contribution is supported by NSERC of Canada. 
UW Madison acknowledges support by NSF award PHY-1903316.
The AWAKE collaboration acknowledges the SPS team for their excellent proton delivery.
\end{acknowledgments}

\bibliography{main_text}

\begin{thebibliography}{40}
\bibitem [{ref()}]{refomegape}%
  \BibitemOpen
  \href@noop {} {\ }\bibinfo {note} {$\omega_{pe} =
  \sqrt{\frac{n_{pe}e^{2}}{\epsilon_0 m_e}}$, where $n_{pe}$ is the plasma
  density, $e$ -- the electron charge, $\epsilon_0$ -- the vacuum permittivity,
  $m_e$ -- the electron mass.}\BibitemShut {Stop}%
\end{thebibliography}


\end{document}